\documentclass[12pt]{iopart}
\usepackage{graphicx,amssymb,amsthm,euscript,setstack}

\newcommand{\ssy}[6]{#1 #4  {\it #2} #6  {\bf #3} #5\rlap{.}}

\newcommand{\numlabel}[1]{\addtocounter{eqnval}{-1}
\refstepcounter{eqnval}\label{#1}}

\newcommand*{\ogr}[1]{\, \rule[-0.76em]{0.4pt}{1.4em} \,
\raisebox{-.55em}{$\displaystyle{}_{#1}$}  }%

\DeclareSymbolFont{iso}{U}{txmia}{m}{it}
\DeclareMathSymbol{\rea}{\mathalpha}{iso}{"92}
\DeclareMathSymbol{\eucl}{\mathalpha}{iso}{"85}
\date{}
\begin{document}
\title{Electrostatic interaction of a pointlike charge with a wormhole}
\author{S Krasnikov}
\address{The Central Astronomical Observatory of Russian Academy of
Science, M-140, Pul\-ko\-vo, St.~Petersburg, Russia}
\ead{Gennady.Krasnikov@pobox.spbu.ru}

\begin{abstract}
A pointlike static (or quasistatically moving) electric charge $q$ is
considered  in the spacetime which is a  wormhole connecting two
otherwise Minkowskian spaces. The electrostatic force acting on the
charge is found to be a sum of two terms. One of them is uniquely
determined by the value of $q$ and the geometry of the wormhole.  The
other has the Coulomb form and is proportional to a freely specifiable
parameter (the ``charge of the wormhole"). These terms are interpreted,
respectively, as the self-force and the force exerted on the charge by
the wormhole. The self-force is found explicitly in the limit of
vanishing throat length. The result differs from that obtained recently
by Khusnutdinov and Bakhmatov.
\end{abstract}
\section{Introduction}
\begin{figure}[t,b]\begin{center}
\includegraphics[width=0.8\textwidth]{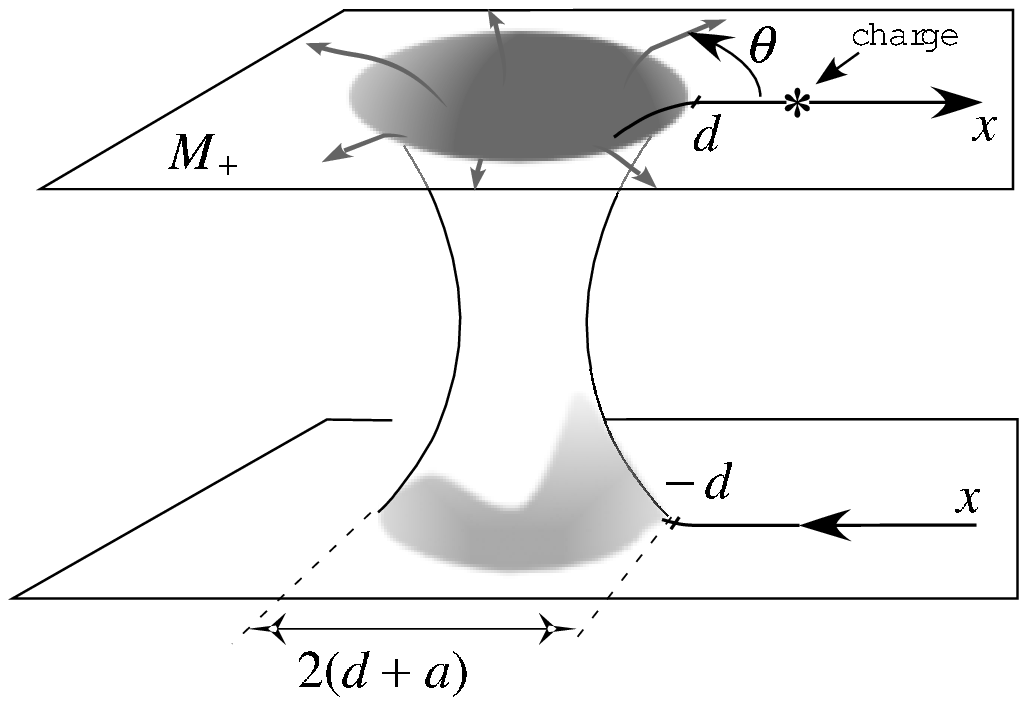}
\end{center}\caption{A section $\varphi=const$ of
\label{fig:kar1}$M^{(3)}$. The divergent gray lines are the force lines
of a source-free field which ``imitates" the Coulomb field for an
observer outside the sphere $r=d+a$.}\end{figure}

What electric force (if any) acts on a pointlike charge  at rest outside
a wormhole, if there are no more charges in the space? This question is
of interest by, at least, two reasons. The first is its relation to the
famous concept of ``charge without charge" \cite{Wheeler}.  Suppose, in a
flat region of a spacetime we observe the electric field
\begin{equation}\label{eq:Coulomb}
\bi
E= Q{}\bi r/r^3 ,\qquad r>r_0.
\end{equation}
 From this we need not conclude that the field is generated by a charge
 (sitting, say, at  $r=0$). It may well happen that there is a
wormhole mouth inside the sphere $r=r_0$ (so that the ``coordinate" $r$
does not, in fact, take the zero value) and the field force lines do not
terminate at all, see figure~\ref{fig:kar1}.
 Which suggests that
maybe there is no charge --- as ``substance" | in nature and the
electromagnetic field is described by the source-free Maxwell equations,
while all elementary ``charges" are, in fact,  mouths of
wormholes\footnote{Throughout the paper we regard the matter supporting
the wormhole as electrically neutral.}. In developing such a theory it
would be important to know how strong the resemblance is between a
wormhole threaded by force lines and a pointlike charge $Q$. The flux
conservation together with the spherical symmetry guarantees that
\eref{eq:Coulomb}  is valid in both cases. So, one might think
that, as long as we restrict ourselves to the region $r>r_0$, the
resemblance is perfect. As we shall see, however,
\emph{this is not the case}: the force $\bi F$ experienced by a
finite charge $q$ put in a point $p_*$ outside the wormhole would
\emph{not} be just the Coulomb force $\bi F_\mathrm{C}=
{qQ}{}\bi r(p_*)/r^3(p_*)$
(moreover, $\bi F_\mathrm{C}$ may turn out to be a negligible
part of   $\bi F$).

Another reason of interest in finding  $\bi F$ is the possible existence
of macroscopic traversable wormholes. At the moment the only
observational restriction on their abundance has been obtained on the
basis of unusual lensing properties of negative mass
\cite{OBS}
 and is valid only for a very
special type of wormholes. To improve the situation it would be
desirable, of course, to know more about physical effects involving
wormholes. The consideration of the electrostatic problem in the wormhole
background can be viewed as a step in that direction. An interesting, in
this sense, result of this paper is that self-interaction leads to
appearance of the attraction infinitely growing (in the approximation of
infinitely short throat) as the charge approaches a mouth of a wormhole.
Tempted by the resemblance between the electrostatic and (Newton's)
gravitational forces | which differ in the \emph{sign}, though | one
might speculate therefore that wormholes are possible which are
macroscopic and static, but nevertheless non-traversable for massive
bodies.

Our analysis will be confined to a simplest wormhole:
\numparts
\begin{equation}\label{eq:metric}
\rmd s^2 = -\rmd t^2 + \rmd  x^2 +r^2( x)(\rmd \theta^2 + \sin^2 \theta
\,\rmd \varphi^2)
\end{equation}
\begin{equation}
\label{eq:usl na r}
 x\in\rea,\qquad r\in C^\infty,\quad r(-x)=r(x),\quad r>0,\quad
 r\ogr{x>d}=x+ a.
\end{equation}
\numlabel{eq:WH}\endnumparts
The wormhole is obviously static and spherically symmetric. Each its
spacelike section $t=const$ --- we shall denote such sections by $
M^{(3)}$
--- is a pair of flat three-dimensional spaces $M_+$ and $M_-$ (they are
defined by the inequalities $x>d$ and $x<-d$, respectively, and either is
just the Euclidean space minus a ball of radius $d+a$) connected with a
`tunnel', see figure~\ref{fig:kar1}.


It is well known
\cite[problem 14.16]{lppt} that in curved spacetime the Maxwell equations
written for the vector-potential $A^i$ have, in the general case, two
non-equivalent versions. Fortunately,  $R^0_i=0$ for our metric and the
difference does not lead to any ambiguity in the equation on $\Phi=A^0$.
It reads:
\numparts
\begin{equation}\label{eq:fund'}
  \Phi,_{a}{}^{;a}(p,p_*)=-4\pi q\delta(p -  p_*),\qquad p\in M^{(3)},
\end{equation}
where $a=1,2,3$ and the derivatives   are by the coordinates of $p$ (not
$p_*$). Equation~\eref{eq:fund'} can be solved by standard methods, see
the following section, but there are two problems in finding the force
$\bi F$ from $\Phi$:
\paragraph{A.} The solutions of \eref{eq:fund'}
diverge in $p_*$, where the force is to be found, and thus one needs a
``renormalization" procedure to derive a meaningful and finite value for
the force. The problem is quite hard in the general case, see, e.~g.,
\cite{quinn-wald-97,khus}, and references therein, but in the
case under discussion  $p_*$ is restricted to $ M_+$, where the procedure
is trivial due to  flatness: to obtain the renormalized solution
$\Phi^\mathrm{ren}$ one simply subtracts the Coulomb part from $\Phi$,
see
\cite{quinn-wald-97,linetSF}.

\paragraph{B.} The  more serious problem is that \eref{eq:fund'} has too many
solutions:  if some $\Phi_1$ solves~\eref{eq:fund'} then so also does
\[
\Phi_1(p,p_*) + f(p_*)Z(p),
\]
where $f$ is arbitrary and the ``source-free" potential $Z$ is an
arbitrary harmonic function.
In the ordinary electrostatics the problem is
solved by requiring  the electric field to fall at infinity
\begin{equation}\label{eq:pot-fall}
  \Phi,_i \to 0 \qquad \mathrm{ at}\quad r_*\equiv r(p_*)=const,\quad
  r(p)\to\infty,
\end{equation}
\numlabel{eq: na pot-l}
\endnumparts
which physically means that we are not interested in field
configurations with infinite energies. We adopt the restriction
\eref{eq:pot-fall} too, but in our case this does not fix the problem,
because in $M^{(3)}$ there \emph{are} non-zero harmonic functions
satisfying~\eref{eq:pot-fall}. Thus, the (absolute value of) the force
experienced by  a pointlike charge near the wormhole is \emph{arbitrary}
and the question posed in the beginning of the paper has no meaningful
answer. To overcome this problem I introduce ``the charge" of the
wormhole defined | up to the factor $4\pi$ | as the flux of $\bi E$
through the throat and prove (see the proposition in the following
section) that\emph{ for the wormhole of a given charge $Q$ the solution
of
\eref{eq: na pot-l} is unique} up to an additive constant. The solution
depends on $Q$ in quite a natural way:
$$
\Phi^\mathrm{ren}(q,p,p_*)= Q/r(p) + \Phi_\mathrm{sf}(q,p,p_*)+ const.\eqno(*)
$$
[cf.~\eref{eqs:Phi_ext}]. The desired force  acting on the charge is
found now by, first, differentiating this expression  by the coordinates
of $p$ and then setting $p=p_*$:
\numparts
\begin{equation}\label{eq:arbQ}
\bi F(r)=  {qQ(r)}{}\bi r/r^3-q\nabla\Phi_\mathrm{sf}(r,r)
\end{equation}
(as before, $\nabla$ in this expression acts on the first argument).

Formally, equation \eref{eq:arbQ} solves the problem in discussion [an
explicit expression for $\Phi_\mathrm{sf}$ is given by
\eref{eqs:Phi_ext}]. It is, however, of  little practical use yet.
Indeed, by $\bi F(p)$ one normally understands the dependence of the
force on position of the charge when everything except the position is
assumed to be fixed. But this latter (perhaps, somewhat vague) condition
in no way enters the derivation of \eref{eq:arbQ} and the function $Q(r)$
is therefore \emph{arbitrary}. To fix it suppose that the charge is
transported (quasistatically, so that the radiation can be neglected)
from $p_*$ to some other point $p_{**}\in M_+$. In section~\ref{sec:S-i},
I argue that $Q$ in such a case will remain unchanged $Q(r_*)=Q(r_{**})$,
which conservation justifies the  name ``charge". So, \eref{eq:arbQ} must
be complemented with
\begin{equation}\label{eq:Qconst}
  Q(r)=const,
\end{equation}
which accomplishes the task.
\numlabel{complF}
\endnumparts

The structure of \eref{complF} with $Q$ independent of $p_*$ and $
\Phi_\mathrm{sf}$ independent of $Q$ suggests interpretation of the first
term in \eref{eq:arbQ} as the force exerted on the charge by the
source-free field, or by the wormhole. And the second term is naturally
interpreted as the self-force.

\paragraph{Note.} Recently, Khusnutdinov and Bakhmatov \cite{khus} have
found special solutions --- let us denote them by
$\Phi_\mathrm{KB}^{(1)}$ and $\Phi_\mathrm{KB}^{(2)}$ --- of
equations~\eref{eq: na pot-l} for $r =
\sqrt{a^2+x^2}$ and $r = a + |x|$, respectively (later
$\Phi_\mathrm{KB}^{(1)}$ was refound by Linet \cite{linet}, who used a
different method). Neither of those $r(x)$ satisfies \eref{eq:usl na r},
but the main problem with finding the self-force [which is how to
identify the self-interaction potential among the infinitely many
solutions of \eref{eq: na pot-l}] is the same as in our case.
Correspondingly, as explained above, the quantity
$-q\nabla\Phi_\mathrm{KB}^{\mathrm{ren}(i)}(r,r)$ need not be the
self-force. And, indeed, calculating the flux of $\nabla
\Phi^{\mathrm{ren}(1)}_\mathrm{KB}$ through the sphere $x=const$  one finds that it
\emph{depends on $p_*$}, see, e.~g., \cite[(20)]{linet} and the sentence
below it. Likewise, for the wormhole of the second type the comparison of
our formula
\eref{eq:Phi_ext} with that for $G^{ren}$ in
\cite{khus} gives the flux  $-qa/(2r_*)$. So,
$\mathbf F\equiv-q\nabla\Phi^{\mathrm{ren}(i)}_\mathrm{KB}(p_*,p_*)$ is
\emph{not} the self-force, but rather another, much less meaningful,
quantity | the force acting on the pointlike charge located in $p_*$ in
the presence of a wormhole with the charge $Q(p_*)$.

\section{The multipole expansion}
\label{sec:gen}

In this section we establish the uniqueness of the  solution of
equation~\eref{eq: na pot-l}  up to the term $Q\rho/r +
\Phi_0$, where $\rho$ is a
certain function of $r$ (specified below), while $Q$ and $\Phi_0$ do not
depend on $r$.

We begin by rewriting equation~\eref{eq:fund'} in the coordinate form
\begin{eqnarray*}
\Big[
\partial_ x^2+\frac{2r'}{r}\partial_ x +\frac{1}{r^2}(\partial_\theta^2 +
\cot  \theta\,\partial_\theta +\sin^{-2}\theta\,\partial_\varphi^2)\Big]
&\Phi
\\&=
-\frac{4\pi q
}{r^2\sin\theta}\,\delta(\varphi)\delta(\theta)\delta(x-x_*)
\end{eqnarray*}
(we have set $\varphi_*=\theta_*=0$, which obviously does not lead to any
loss of generality). Expanding
\[
 \Phi=\sum_{l=0}^\infty\sum_{m=-l}^{l}\phi_l^{(m)}(x)Y_l^m(\varphi,\theta),
\]
where $Y_l^m$ are spherical functions \cite{BT}
\[
Y_l^m(\varphi,\theta)\equiv
\sqrt{\frac{2l+1}{4\pi}\,\frac{(l-|m|)!}{(l+|m|)!}}\mathcal
P_{l}^{|m|}(\cos\theta)  e^{im\varphi},
\]
\[
\mathcal P_{l}^m(\mu)\equiv(1-\mu^2)^{\textstyle\frac m2}
\:\frac{\rmd^m }{\rmd\mu^m}\,\mathcal P_{l}(\mu)
\]
($\mathcal P_{l}$ are the Legendre polynomials) one gets
\begin{eqnarray}
 \sum_{l=0}^\infty\sum_{m=-l}^{l}\Big[\partial_ x^2+\frac{2r'}{r}
\partial_x - \frac{l(l+1)}{r^2}\Big]\phi_l^{(m)}(x)&Y_l^m(\varphi,\theta)
\nonumber\\
&=\label{eq:expan}
 -\frac{4\pi q
}{r^2\sin\theta}\,\delta(\varphi)\delta(\theta)\delta(x-x_*).
\end{eqnarray}
Multiply both sides of \eref{eq:expan} by $r{Y_{l'}^{m'*}}\sin\theta$ and
integrate over $\varphi$ and $\theta$. The result ($Y_{l'}^{m'}$ are
orthonormal on the sphere) is
\[
\Big[\partial_x^2- \Big(\frac{r''}{r}+
\frac{l(l+1)}{r^2}\Big) \Big] r\phi_l^{(m)}(x)
=
-\frac{4\pi q}{r}\,\delta(x-x_*)Y_{l}^{m*}(0,0),
\]
It is convenient to treat the cases of zero and non-zero $m$ separately,
because
\[
Y_{l}^{0}(0,0)=\sqrt{\frac{2l+1}{4\pi}},
\qquad
Y_{l}^{m}(0,0)=0, \quad m\neq 0
\]
So, we define
\[
 v_l\equiv \sqrt{ \textstyle{\frac{2l+1}{4\pi}} } r\phi_l^{(0)},\qquad v_{l,m}\equiv
 r\phi_l^{(m)},\quad m\neq 0.
\]
For $v_l$ we have
\begin{equation}\label{eq:ur na phi}
\Big[\partial_x^2- \Big(\frac{r''}{r}+ \frac{l(l+1)}{r^2}\Big) \Big]
v_l= -\frac{2l+1}{r_*}\,q\delta(x-x_*)
\end{equation}
while $v_{l,m}$ irrespective of $m$ must solve the equation
\begin{equation}\label{eq:ur na psi}
\Big[\partial_x^2- \Big(\frac{r''}{r}+ \frac{l(l+1)}{r^2}\Big) \Big]
z(l,x) =0.
\end{equation}
Thus, the solution of \eref{eq: na pot-l} is the function
\begin{equation}\label{eq:Phi'}
\Phi=\frac{1}{r} \,\sum_{l=0}^\infty v_l(x)\mathcal
P_{l}(\cos\theta)  +
\frac1r\, \sum_{l=1}^\infty\sum_{|m|=1}^lv_{l,m}(x)Y_l^m(\varphi,\theta),
\end{equation}
where $v_l$ and $v_{l,m}$ are the solutions, respectively, of
\eref{eq:ur na phi} and \eref{eq:ur na psi} which [because of
\eref{eq:pot-fall}]  grow at $|x|\to\infty$ not faster than $|x|$.

To proceed note that in the flat regions $M_\pm$  the term with $r''$
vanishes in \eref{eq:ur na psi} and the equation is easily solved: the
solution is a superposition of $r^{-l}$ and $r^{l+1}$.
\paragraph{Notation}
By $z_-$ and $z_+$ we denote the solutions of \eref{eq:ur na psi} which
are equal  to $r^{-l}$ at, respectively,  $x<-d$ and $x>d$. And $z_ e$,
$z_ o$ are the solutions of \eref{eq:ur na psi} defined by the initial
data
\[
z_ e(0)=1,\quad z'_e(0)=0,\qquad z_ o(0)=0,\quad z'_o(0)=1
\]
Evidently  $z_ e$ and $z_ o$ are even and odd, respectively, and any
solution of \eref{eq:ur na psi} is their linear combination.
\paragraph{Proposition.} If $z(l,x)$ is a solution of \eref{eq:ur na psi}
with  $l> 0$, the function $r^{-1}z$ grows unboundedly as $x\to
(-)\infty$.

\begin{proof}
We start with the observation that if a solution $z$ of
\eref{eq:ur na psi} satisfies the condition
\numparts\numlabel{obs-n}
\begin{equation}\label{eq: usl polo}
   z(x_0) >0,\qquad W[z ,r](x_0)\geq 0,
\end{equation}
where $W$ is the Wronskian $W[f_1,f_2]=f_1'f_2 -f_1f_2'$, then
\begin{equation}\label{eq: rost}
z'/z>0\quad \mathrm{and}\quad z/r\mathrm{\ grows\qquad at}\ x>x_0.
\end{equation}
\endnumparts
Indeed, rewrite \eref{eq:ur na psi} as
\begin{equation}\label{eq:Vron}
W'[z ,r]=\frac{l(l+1)}{r}z.
\end{equation}
Integrating this equation one gets
\begin{equation}\label{eq:Vron int}
\frac{z '} {z } - \frac{r'}{r}=
\frac{1}{rz }W[z ,r](x_0)+
\frac{l(l+1)}{rz }\int_{x_0}^x\frac{z \,\rmd x }{r}.
\end{equation}
Due to \eref{eq: usl polo} the r.~h.~s.\ is positive at least up to
$x_1$, where $x_1$ is
$
\infty$, if $z(x)$ has no zeroes, and the first zero of $z$ otherwise. Thus,
$z'/z>0$ and $z/ r $ grows at $ x\in (x_0,x_1)$. The latter means, in
particular, that $x_1$ cannot be finite (because if it were,  $r(x_1)$
would have been less than  $z(x_1)=0$), which  proves \eref{eq: rost}.

Now note that both $z=z_ {e}$ and $z=z_ {o}$ satisfy \eref{eq: usl polo}
with $x_0$ equal to zero in the former case and to some (sufficiently
small) positive number in the latter. So, $z_ {e(o)}/ r $ grows at all
$x>x_0$ and hence, $z_ {e(o)}$ cannot be proportional to $r^{-l}$ at
large $x$. Consequently,
\begin{equation*}
z_ {e(o)}(x) \sim c_2r^{l+1},\qquad x\to\infty,\quad c_2\neq 0.
\end{equation*}
The same is true for $x\to-\infty$, since $z_ {e(o)}$ is even (odd). And,
finally, it is true, when  $x\to\infty$ or $x\to-\infty$, for
\emph{every} $z$ because any of them is a superposition of $z=z_ {e}$ and
$z=z_ {o}$.
\end{proof}

\paragraph{Corollary 1.}  If  $l> 0$, the solutions  $z_+$ and $z_-$ are
linearly independent.

\paragraph{Corollary 2.}  The second term in the r.~h.~s.\ of
\eref{eq:Phi'} is zero.

\paragraph{Corollary  3.} Denote by $\vartheta$ the Heaviside step function.
Then at $l> 0$ the function
\numparts
\begin{equation}\label{eq:poln v_n}
v_l=-\frac{(2l+1)q }{ r_*}\,
\frac{ \vartheta(x-x_*)
z_+(l,x)z_-(l,x_*) +
\vartheta(x_*-x)z_-(l,x)z_+(l,x_*)}{W[z_+,z_-]}
\end{equation}
is the unique solution of \eref{eq:ur na phi} that  grows slower than
$r^{l+1}$ at the infinities.

Now let us turn to the case $l=0$. Equation \eref{eq:ur na psi} [as seen
from
\eref{eq:Vron}] transforms into $W'[ z,r] =0$. This gives
\[
 (z/r)'=C/r^2,
\]
where $C$ is an arbitrary constant. Thus, $z(0,x)$ is a linear
combination of $r$ and~$\rho$
\[
\rho(x)\equiv \frac{r(x)}{r(d)}- r(x)\int_d^ x\frac{\rmd x}{r^2(x)} .
\]
In this case $z_+$ is proportional to $z_-$ and the formula
\eref{eq:poln v_n} does not define $v_0$. The latter, however, can be
easily found by using --- as independent solutions of the homogeneous
equation \eref{eq:ur na psi} --- the functions $\rho$ and $r$ instead of
$z_+$ and $z_-$ (note that $W[ r,\rho]=1$):
\begin{equation}\label{eq:poln v_0}
v_0=\frac{q }{ r_*}\,\Big( \vartheta(x-x_*) \rho(x)r_* +
\vartheta(x_*-x)r\rho_*\Big) + Q\rho + \Phi_0r
\end{equation}
Here $\rho_*=\rho(x_*)$  and
 $Q$, $\Phi_0$ are arbitrary, but do not depend on $r$.
\numlabel{eq:poln v}
\endnumparts

Summing up,
\begin{equation}\label{eq:okonch}
\Phi=\frac1r \,\sum_{l=0}^\infty v_l(x)\mathcal
P_{l}(\cos\theta),
\end{equation}
where $ v_l(x)$ are given by formulae \eref{eq:poln v}.

\section{Self-interaction}\label{sec:S-i}
Equation~\eref{eq:okonch} gives, in principle, the electrostatic field of
a pointlike charge in the wormhole background. However, as mentioned in
the Introduction, to find the
\emph{force} acting on the charge it remains to cope with the fact that
the field diverges in the point $p_*$ where the charge is located. To
this end we take advantage of the fact that $p_*$ is in a flat part of
the wormhole (let it be $M_+$, for definiteness). In this region we
define the potential (for the second equality see, e.~g.,
\cite[(II 2.13)]{BT})
\begin{eqnarray}
 \Phi_\mathrm{Eucl}(p,p_*)\equiv \frac{q}{ |p,p_*|}
=
\nonumber \\\label{eq: Phi_ch}
\frac 1r\sum_{l=0}^\infty
q[\vartheta(x-x_*)(r_*/r)^{l} + \vartheta(x_*-x)(r/r_*)^{l+1}]\mathcal
P_{l}(\cos\theta) ,
 \end{eqnarray}
where $ |p,p_*|$ is the distance between $p$ and $p_*$ in the space
$\eucl^3$, obtained by gluing a usual Euclidean ball of radius $d$ to
$M_+$. From the usual electrostatics we know that the field
$-\nabla\Phi_\mathrm{Eucl}$ exerts no force on the charge. So, in finding
the self-force we are only interested in the difference
\begin{equation}\label{eq: Phi_wh}
\Phi^\mathrm{ren}\equiv \Phi -\Phi_\mathrm{Eucl}
\end{equation}
(which is defined, of course, only in $M_+$). It is $\Phi^\mathrm{ren}$
that plays the r\^ole of the ``external field", i.~e.\ the force acting
on the charge is $\bi F=-q\nabla \Phi^\mathrm{ren}(p_*,p_*)$.

To rewrite the expression \eref{eq:okonch} for $ \Phi$ in a more
convenient form let us substitute the equalities (in fact, the  second
one is a definition of  $\alpha_l $)
\begin{equation}\label{eq:def alpha}
z_+(x,l)=r^{-l}, \quad z_-(x,l)=C(r^{l+1} + \alpha_l r^{-l}),\qquad
\mathrm{at}\quad l> 0,\ x>d,
\end{equation}
 into
\eref{eq:poln v_n}:
\begin{eqnarray*}   v_l=
q[\vartheta(x-x_*)(r_*/r)^{l} + \vartheta(x_*-x)(r/r_*)^{l+1}
+\frac{\alpha_l}{r_*}(rr_*)^{-l}],\\\mathrm{at}\quad l> 0,\quad x,x_*>d
 \end{eqnarray*}
Substituting  this together with an obvious (notice that $\rho(x)=1$ at
$x>d$) equality
\begin{equation*}
v_0=q[\vartheta(x-x_*) + (r/r_*)\vartheta(x_*-x)] +
 Q + r\Phi_0,\qquad \mathrm{at}\quad x,x_*>d.
\end{equation*}
into \eref{eq:okonch} and, then, the result | combined with
\eref{eq: Phi_ch} | into \eref{eq: Phi_wh}, we finally obtain
\numparts
\begin{equation}
\Phi^\mathrm{ren}(p,p_*)=\Phi_\mathrm{sf}(p,p_*)+\Phi_\mathrm{wh}(p,p_*),\qquad
p,p_*\in M_+,
\end{equation}
where
\begin{equation}\label{eq:Pole kr}
\Phi_\mathrm{sf}\equiv q\sum_{l=1}^\infty \alpha_l(rr_*)^{-l-1}\mathcal
P_{l}(\cos\theta),
\qquad \Phi_\mathrm{wh}\equiv  Q/r + \Phi_0.
\end{equation}
\numlabel{eqs:Phi_ext}
\endnumparts
\paragraph{Note} In the region under consideration (i.~e., at $x,x_*>d$)
$\Phi_\mathrm{sf}$ is smooth.
\begin{proof}
By definition [see, \eref{eq:def alpha}]
\begin{eqnarray}
\mathrm{at}\quad x=d\qquad
\alpha_l=-\frac{(z_-r^{-l-1})'r^{2(l+1)}}{(2l+1)C} =\nonumber
\\ \label{eq: all}
-\frac{(z_-r^{-l-1})'r^{2(l+1)}}{(2l+1)}
\;\frac{(2l+1)r^{2l }}{(z_-r^{l})'}
=\frac{ l/r -z'_-/z_- +1/r}{l/r  + z'_-/z_-  }\,r^{2l+1}(d)
\end{eqnarray}
On the other hand, $z_-$ satisfies the condition \eref{eq: usl polo} with
$x_0=-d$. Hence, by \eref{eq: rost}, $z'_-/z_-$ is positive at $x=d$.  It
follows then from \eref{eq: all} that at $l\to\infty$
\[
\alpha_l = A_lr^{2l+1}(d), \qquad A_l=O(1)
\]
and
\[
\Phi_\mathrm{sf}(x,x_*)= q\sum_{l=1}^\infty \frac{A_l}{d+a}
\Big[\frac{r(d)}{r(x)}\,\frac{r(d)}{r(x_*)}\Big]^{l+1}\mathcal
P_{l}(\cos\theta).
\]
Obviously for  any $x_1>d$ the series converges uniformly on
$[x_1,\infty)$ and so do all the series obtained from this one by
termwise differentiation in $r$.
\end{proof}

We interpret $\Phi_\mathrm{sf}$ and $\Phi_\mathrm{wh}$ as the parts of
$\Phi^\mathrm{ren}$ generated by the charge and by the wormhole,
respectively. To justify this interpretation  note that 1)
$\Phi_\mathrm{sf}$, for a given $p$, depends only on $q$ and $p_*$ and 2)
$\Phi_\mathrm{wh}$, in contrast, does not depend on $p_*$ in the
following sense. Suppose the charge is at rest up to some moment $t_0$
and is then quasistatically moved from $p_*(t_0)$ to some $p_{*}(t_1)\in
M_+$, where | at the moment $t_1$ | is again put to rest. Let $t_2$ be a
moment when at small $|x|$ the disturbance in the potential caused by the
motion of the charge has already settled down and the potential became
constant (in time)\footnote{The existence of such a moment is an
\emph{assumption}, even though a very plausible one. If a wormhole is
such that in its vicinity the electro-magnetic waves caused by stirring
the charge do not dissipate with time, one probably cannot develop
electrostatics in that spacetime at all.}. Then at times $t>t_2$ in the
vicinity of the wormhole
 the equations~\eref{eqs:Phi_ext} remain valid with $r_*(t_0)$ replaced
by $r_{*}(t_1)$ and with the \emph{same} $Q$.
\begin{proof}
Indeed, at $t>t_2$ the flux of $\nabla\Phi$ through the sphere $x=d$ is
$\EuScript F(d,t_2)=-4\pi Q(t_2)$, because neither $\Phi_\mathrm{Eucl}$,
nor $\Phi_\mathrm{sf}$ give any contribution to it. At the same time,
there is a sphere $x=D>x_*$ such that $\EuScript F(D,t_2)=-4\pi
[Q(t_0)+q]$, because if $D>x_*+ c(t-t_0)$, the field is not disturbed
there yet. Thus the flux $\EuScript F_B(t_2)$ through the boundary of the
layer $\{d<x<D, t=t_2\}$ is $4\pi[Q(t_2)-Q(t_0)-q]$. On the other hand,
the total charge inside the layer has not changed and hence $\EuScript
F_B(t_2)=\EuScript F_B(t_0)=-4\pi q$ by the Gauss theorem. So,
$Q(t_2)=Q(t_0)$.
\end{proof}

\section{Short wormhole}
It is seen from formulae \eref{eqs:Phi_ext} that the force acting on a
charge depends on the form of the wormhole | the information about the
form being encoded in the coefficients $\alpha_l$. But today we have no
reason to consider any particular form as more realistic than any other.
So, it would be interesting to find a form-independent effect. To this
end we consider in this section the limit $d\to 0$ for the
wormhole~\eref{eq:WH} with $a>0$. In doing so we allow the the throat to
be \emph{arbitrary}, the only additional requirement on $r(x)$ being
\begin{equation}\label{eq:treb}
 r'< c_r\qquad\forall d
\end{equation}
($c_r$ is a constant), which, among other things, guarantees that $r(0)
\to a$.

Let us, first, present $z_-$ as the solution of the following
differential equation (which does not contain the large quantity $r''$)
\begin{eqnarray*}
\Big[\partial_ x^2+2r'r^{-1}
\partial_x - l(l+1)r^{-2}\Big]y=0,
\\
 y(-d)=(d+a)^{-l-1}.
\end{eqnarray*}
Here the first line is simply \eref{eq:ur na psi} in terms of $y\equiv
z_-/r$, while the second
 follows from the definition of $z_-$. The
coefficients of the equation by
\eref{eq:treb} are uniformly (by $d$) bounded, so at a \emph{fixed} $l$
and $d\to 0$
\[
\ln' y(-d) \to \ln' y(d).
\]
Hence, $\ln' z(-d) -  \ln' r(-d)\to \ln' z(d) -  \ln' r(d)$ and thus
\[
\ln' z_-(d) \to \ln' z_-(-d) + 2/a \to  (l+2)/a
\]
(recall that $z_-=(a-x)^{-l}$ at $x=-d$). On the other hand, by
\eref{eq:def alpha}
\[
\ln' z_-(d)=\frac{(l+1)r^{l} - l\alpha_l r^{-l-1} }{r^{l+1} + \alpha_l
r^{-l}}\ogr{r=a}=\frac{(l+1)a^{l} - l\alpha_l a^{-l-1} }{a^{l+1} +
\alpha_l a^{-l}},
\]
combining which with the equation above we find in the limit $d\to 0$
\[
\alpha_l= -\frac{a^{2l+1}}{2(l+1)}.
\]
Thus, asymptotically,
\begin{eqnarray*}  \Phi^\mathrm{ren}\sim -q\sum_{l=1}^\infty
\frac{a^{2l+1}}{2(l+1)}&(rr_*)^{-l-1} \mathcal P_{l}(\cos\theta)
+ \frac Qr +\Phi_0
\\
&=-\frac q{2a}\sum_{l=2}^\infty
\frac{1}{l}(a^{2}/rr_*)^{l}\mathcal P_{l-1}(\cos\theta) +
\frac Qr + \Phi_0.
 \end{eqnarray*}
In particular, at $\theta=0$
\begin{eqnarray}
 \Phi^\mathrm{ren}\sim -\frac q{2a}\sum_{l=2}^\infty
\frac{1}{l}(a^{2}/rr_*)^{l}& +
\frac Qr + \Phi_0
\nonumber
\\\label{eq:Phi_ext}
&=\frac q{2a}\left[\ln(1 - \frac{a^{2}}{rr_*}) +\frac{a^2}{r_*r}\right] +
\frac{ Q}{r} + \Phi_0
\end{eqnarray}
and the  electric field on the axis  is
$$
-\Phi,_x=-(\Phi^\mathrm{ren}+\Phi_\mathrm{Eucl}),_x\sim q\frac{
r-r_*}{|r-r_*|^3}\,+\frac{qa^3}{2r_*r^2(a^2-rr_*)}+\frac{ Q}{r^2}.
$$

Thus, asymptotically, in the presence of an infinitely short wormhole
with the radius $a$ a pointlike charge $q$ experiences the (radial) force
\[
F(r_*)=-q\Phi^\mathrm{ren}_{,x}=-\frac{q^2a^3}{2r_*^3(r_*^2-a^2)}+\frac{
qQ}{r_*^2}.
\]
Its first term | the self-force $\bi F_\mathrm{s}$ | can be presented, if
desired, in the form
\[
\bi F_\mathrm{s}(r_*)= -\nabla U(r_*), \qquad U(r_*)=\frac{q^2}{4a}
\Big(\ln [1 - (a/r_*)^2 ] +(a/r_*)^2.
\Big)
\]

\section*{Acknowledgements}
I am grateful to N. R. Khusnutdinov and R. R. Zapatrin for an inspiring
discussion on the subject.  This work was supported by RNP Grant
No.~2.1.1.6826.

\end{document}